\documentclass[9pt,preprint2,natbib209]{aastex61}

\usepackage{color}
\usepackage{graphicx}
\usepackage{dcolumn}
\usepackage{bm}

\def\apjs{ApJS}

\def\apjl{ApJL}

\def\aap{A\&A}

\begin{document}

\title[The Origin of Isolated Millisecond Pulsars]{Quark Novae: An Alternative Channel For the Formation of Isolated Millisecond Pulsars}

\correspondingauthor{Chunhua Zhu}\email{chunhuazhu@sina.cn}
\author{Nurimangul Nurmamat}\email{nurimangul\_uy@sina.com}
\author{Chunhua Zhu}
\affil{School of Physical Science and Technology,
Xinjiang University, Urumqi, 830046, China}

\author{Guoliang L\"{u}}
\affil{School of Physical Science and Technology,
Xinjiang University, Urumqi, 830046, China}

\author{Zhaojun Wang}
\affil{School of Physical Science and Technology,
Xinjiang University, Urumqi, 830046, China}

\author{Lin LI}
\affil{School of Physical Science and Technology,
Xinjiang University, Urumqi, 830046, China}

\author{Helei Liu}
\affil{School of Physical Science and Technology,
Xinjiang University, Urumqi, 830046, China}

\begin{abstract}
Isolated millisecond pulsars (IMSPs) are a topic of academic contention.
There are various models to explain their formation. We explore the formation of IMSP via quark novae (QN).
During this formation process, low-mass X-ray
binaries (LMXBs) are disrupted when the mass of the neutron star (NS) reaches 1.8 $\rm M_\odot$.
Using population synthesis, this work estimates that the Galactic birthrate of QN-produced IMSPs lies between $\sim 9.5\times10^{-6}$ and $\sim1.7\times10^{-4}$ $\rm yr^{-1}$. The uncertainties shown in our experiment model is due to the
QN's kick velocity. Furthermore, our findings not only show that QN-produced IMSPs are statistically more significant than those produced by mergers, but also that millisecond pulsar binaries with a high eccentricity may originate from LMXBs that have been involved in, yet not disrupted by, a QN.
\end{abstract}

\keywords{pulsars: general- stars: neutron- binaries: close}
\section{INTRODUCTION}
 Millisecond Pulsars (MSPs) are a particular type of pulsar, well-known for their old age.
 They differ from normal pulsars, not only due to their short spin period ($\leq 30$ ms)
but also because of their low magnetic field ($\sim10^{8}$G)\citep[e.g.,][]{Lorimer1996,Tauris2016}.
In most cases, normal pulsars have a long slow-down rates $(\dot{P}\sim10^{-15})$,
whereas the slow-down rate of MSPs is usually $\rm \dot {P}\sim10^{-19}$.
MSPs¡¯ stable rotations and sharp radio pulses make them exceptionally stable clocks \citep{Rawley1987,Backer1987},
being used by the International Pulsar Timing Array Project to detect gravitational waves\citep{Manchester2013}.

MSPs are usually formed in low- and intermediate-mass X-ray binaries through a process of recycling \citep{Alpar1982}, whereby Roche lobe overflow from its companion star causes a neutron star (NS) to not only accrete mass $(\geq0.1\rm M_\odot)$ and angular momentum over a lengthy period of time $(\sim10^{8}\rm yr)$, but also increase in rotational speed up to the millisecond \citep{Alpar1982,Bhattacharya1991,vandenHeuvel2008}.
This may lead one to surmise that the majority of MPSs have evolved from and exist solely in binary systems. However, observational data from the Pulsar Catalogue at ATNF (Australia Telescope National Facility) shows a total of 360 MSPs, approximately 120 of which exist in isolation \citep{Manchester2005}.
These have been called Isolated Millisecond Pulsars (IMSPs) \citep{van1988}.

The origin of IMSPs is a topic of academic contention. Investigations undertaken by \cite{VandenHeuvel1984} have suggested
that a close binary system, not only consisting of a NS and a massive white dwarf (WD), but also with a $\rm P_{\rm orb}<16$ hours, may inescapably merge into an IMSP, as a consequence of gravitational radiation loss $(\rm M>0.7\rm M_\odot)$. For such NSs and massive WDs $($>$0.9\rm M_\odot)$, \cite{Davies2002}  have proposed a formation rate of $5\times10^{-4}\sim5\times10^{-5}$ yr$^{-1}$.
 \cite{Ferrario2007} and \cite{Gr2013}, however, have suggested that the formation rate of IMSPs ranges from $3\times10^{-6}\sim5\times10^{-6}$ yr$^{-1}$. This figure is significantly lower than the Galactic birth rate of IMSP formed through the merging of massive WDs and NSs, hence rendering it somewhat unreliable. What¡¯s more, in recent literature,
\cite{Sun2018} used population synthesis to investigate the possible origin of IMSPs as product of NS and WD coalescence, putting forward the estimate that IMSPs¡¯ Galactic birthrate may lie between $5.8\times10^{-5}$ $\rm yr^{-1}$ and $2.0\times10^{-4}$ $\rm yr^{-1}$.

In \cite{Camilo1993}, it is suggested that IMSPs may be formed after recycled high-mass X-ray binary systems are disrupted by the evolution of high-mass companion stars into supernovae. However, according to data from \cite{Belczynski2009}, only 10\% of disrupted recycled pulsars are IMSPs.
For this reason, one can infer that disrupted recycled pulsars play but a small part in the formation of IMSPs.

\cite{Portegies2011} put forward a triple-star theory to explain the formation of MSPs that are both in binaries with normal stars and have wide orbits with a high eccentricity, such as J1903+0327. Their results indicated that MSPs can be ejected from triple-star systems, due to their dynamic instability, thereby bringing about the formation of IMSPs. Despite this, the amount of IMSPs produced this way seems somewhat insignificant when compared to their overall number.

According to \cite{Kluzniak1988}, loss of mass in a binary system comprised of an MSP and a low-mass companion star may be due to low-frequency electromagnetic radiation, energetic particles, or $\gamma$-rays produced by the pulsar¡¯s fast-spinning magnetic dipole. Building on this theory, \cite{Ruderman1989} then suggested that if such irradiation can effectively  evaporate the companion star, then it is possible that it can be entirely ablated, hence forming an IMSP. However, research by \cite{Stappers1998} \cite{Chen2013} and \cite{Lu2017} have estimated that this would take a significantly long time to occur, rendering it extremely difficult to produce IMSPs within the confines of hubble time.

Here, this paper shall propose an alternative theory for the formation of IMSPs. At first, \cite{Witten1984} acknowledged the existence of compact objects made of strange quark matter, called strange stars (SSs).
\cite{Cheng1996} then noted that these SSs were formed in low-mass X-ray binaries (LMXBs) during an accretion-triggered transition phase, which has been called a quark nova (QN)\citep{Ouyed2002}.
Similar to core collapse supernovae, SSs also gain post-QN kick velocity\citep{Ouyed2002,Ouyed2005,O2011,Ouyed2013,Ouyed2015}.
Thus, LMXBs are disrupted and IMSPs are formed.

In the following sections, this paper shall investigate the possibility of QN-induced IMSP formation.
In \S 2 below, we present our hypothesis and provide some details of our modelling algorithm. This is then followed by a discussion of the results and effects of different parameters in \S 3. Finally, \S 4, presents our main conclusions.

\section{MODELS}
In this work, we employ the rapid binary star evolution code (BSE), which was first explored by \cite{Hurley2002} and then further expanded upon by \cite{Kiel2006}. Up to the present, our research team have used the BSE code to investigate several types of binary systems, including symbiotic stars \citep{L2006,L2009,L2011},
X-ray binaries \citep{L2012,Lu2017,Zhu2017}, accreting MSPs \citep{Zhu2015}, classical nova \citep{Sun2016,Rukeya2017}, runaway stars produced by supernovae \citep{Yisikandeer2016},
and Thorne-Zytkow objects \citep{Hutilukejiang2018}.
In particular,
\cite{Zhu2013} used the BSE code to examine QN-induced SS formation, but did not consider the role of kick velocity.
This paper adopts the model previously presented by
\cite{Zhu2013}, not only taking into consideration three different means of NS production (core-collapse supernovae, evolution-induced collapses and accretion-induced collapses), but also the increasing mass of the accreting NS. Following \cite{Zhu2013}, this work also assumes that a QN occurs when the mass of an NS reaches
$\sim$1.8 $M_\odot$.

\subsection{Kick velocity of QN}
Although it is clear that an SS produced by a QN is subject to kick velocity, our understanding of it is extremely limited.
Based on the kick velocity of core collapse supernovae, one can assume that the distribution of kick velocity for QNs follows the Maxwell distribution:
\begin{equation}\label{}
P(\nu_{\rm k})=\sqrt{\frac{2}{\rm \pi}}\frac{\nu_{\rm k}^{2}}{\sigma_{\rm k}^{3}}e^{-\nu_{\rm k}^{2}/2\sigma_{\rm k}^{2}}.
\end{equation}
Furthermore, as stated in \cite{Ouyed2014}, if a QN explosion is asymmetric, then it can provide a kick velocity of 100 km $\rm s^{-1}$.
In order to facilitate discussion on the effect of kick velocities, this work takes ${\sigma_{\rm k}}$ as 50, 100, 200 and 400 km s$^{-1}$, respectively.

\subsection{Remnants of a QN}
After a QN, the NS becomes a SS, which then gains kick velocity ($\sim$ 100 km s$^{-1}$)
and loses mass ($\sim 10^{-3} \rm M_\odot$)\citep{Ouyed2014}.
This work assumes that the total mass lost in each QN is $10^{-3} \rm M_\odot$.
As seen in \cite{Brandt1995}, certain kick velocities and lost masses can be used to calculate whether an LMXB has been disrupted or has survived post-QN. If the LMXBs is disrupted, then an isolated SS appears. If the spin period of the renascent SS is less than 30ms, then an IMSP is born.

Furthermore, this work assumes that QNs do not alter the spin period of the entities involved. In other words, the spin period of a particular SS is presumed equal to that of its corresponding NS prior to the QN. As shown by the theoretical model of interaction between a rotating, magnetized NS and its surrounding matter
\citep[See:][]{Pringle1972,Illarionov1975,Ghosh1978,Lovelace1995,L2012},
the spin periods of an accreting NSs is dependent on
both its mass-accreting rate and its magnetic field.
After a sufficient mass transfer process has taken place, the spin period of an NS may reach an equilibrium spin period \citep{B1991}.
Based on \cite{Lipunov1988}, this can be given by
\begin{equation}
P_{\rm s}^{\rm eq}=5.72M_{\rm
NS}^{-5/7}\dot{M}_{16}^{-3/7}\mu_{30}^{6/7}\ {\rm s} \label{eq:deq}
\end{equation}
when $\dot{\rm M}<\dot{\rm M}_{\rm Edd}$ and
\begin{equation}
P_{\rm s}^{\rm eq}=1.76\times10^{-1}\mu_{30}^{2/3}M_{\rm NS}^{-2/3}\
{\rm s}
 \label{eq:seq}
\end{equation}
when $\dot{\rm M}\ge\dot{\rm M}_{\rm Edd}$.
Here, $\mu_{30}=\mu/(10^{30}{\rm G\ cm^3})$ is the NS's dipole
magnetic momentum, $\dot{\rm M}_{\rm 16}=\dot{\rm M}_{\rm
NS}/(10^{16}{\rm g/s})$ is its accretion rate and $\dot{\rm M}_{\rm Edd}$ is the Eddington accretion rate.

The magnetic field of an accreting NS has a direct impact on its spin period.
\cite{Shibazaki1989} have suggested that accretion of mass may lead to the decay of an NS¡¯s magnetic field and proposed the following equation:
\begin{equation}\label{3}
 B_{\rm NS}=\frac{B_{0}}{1+\triangle M/m_{\beta}}
\end{equation}
where $\triangle \rm M$ is the accreted mass, $\rm m_{\beta}$ is the mass constant for magnetic field decay, $\rm B_{NS}$ is the field strength at time $\rm t$ and $\rm B_{0}$ is initial field strength of NS. Similar to \cite{Shibazaki1989},  this work takes $\rm m_{\beta} =10^{-4}\rm M_\odot$ and $\rm B_{0}=10^{12}$G.

\section{Result}
This paper uses population synthesis to simulate the $10^{7}$ binary system and, therethrough, investigate the total population of IMSPs produced by a particular QN. Our research team has used such methods widely in the past \citep[See:][]{L2006,L2008,L2009,L2012,A2013}.
This work not only adopts \citet{Miller1979}'s initial mass-function for the mass of primary components, but also a flat distribution of mass ratios
\citep[See:][]{Kraicheva1989,Goldberg1994}.
In addition to this, this work also uses the distribution of separations of $\rm \log a = 5X + 1$, where $\rm X$ is a random
variable uniformly distributed in the range [0,1] and the separation
$\rm a$ is in $\rm R_\odot$.

Our models have shown that IMSPs are a product of the disruption caused by QN on LMXBs. There are, however, some LMXBs that can avoid disruption and continue to survive post-QN. These include MSPs with a high eccentricity, called eMSPs.

\subsection{eMSPs}
Based on data taken from the ATNF pulsar catalogue database,
 there are a total of 25 MSPs existing in eccentric binaries (ecc$>$0.01) with low mass companion stars($\rm M_{2}\leq1M_\odot$)\citep{Manchester2005}.
Out of these 25, a total of 16 MSPs are in globular clusters. Since the high eccentricity of these MSPs is due to interactions with single stars\citep{Rasio1995,Bagchi2009}, this paper focuses investigation on the remaining nine. Table \ref{tab:results} below gives a series of physical parameters for the nine MSP binaries in question. In general, MSPs produced through a recycling process are circular in orbit. The formation of eMSPs, however, is still highly debated.
\cite{Freire2014} have proposed that eMSPs may be formed as a result of the rotationally-delayed, accretion-induced collapse of a massive WD. \cite{Antoniadis2014}, on the other hand, states that the high eccentricity of MSPs may be a result of the dynamical interaction between the binary and the circum binary disk. \cite{Jiang2015} suggests that in LMXBs, the increase in mass of an accreting NS may result in an NS-SS phase transition, which then causes the NS to lose gravitational mass. This provides a satisfactory explanation for the origin of a number of eMSPs, including PSRs J2234+06, J1946+3417 and J1950+2414, that have He WDs as companion stars. In recent literature, however, \cite{Stovall2019} has suggested that the formation of eMSPs may be a result of loss of mass during the proto-WD phase, where there is an unstable burning of shell-hydrogen. The theoretical models used in this paper show that eMSPs originate from LMXBs that have survived post-QN, similar to the theory proposed by \cite{Jiang2015}.In addition to this, our investigation has also brought kick velocity into consideration, which was a factor left unconsidered in previous studies.

If the binary system is undisrupted and continues to exist post-QN, then the additional kick velocity and loss of mass may lead to an orbital expansion. In our simulation model, the post-QN orbital parameters of an eMSP¡¯s binary system are determined by the standard formula given by \cite{Brandt1995}.
As can be seen from Figure \ref{fig:ecc} below, our model can be applied to all eMSPs (when $\sigma_{\rm k}=50$ km s$^{-1}$), whereas other models are unable to explain the formation of MSP binaries with low eccentricity and low-mass companion stars. The reason for this is as follows: The lower the kick velocity, the more likely it is that the binary with low-mass companion stars will survive and the lower the eccentricity of the final surviving binary shall be.
\begin{table}
\begin{center}
\setlength{\abovecaptionskip}{0pt}
\setlength{\belowcaptionskip}{10pt}
\caption{Parameters for the nine observed eMSPS. The first column provides their names, the second column presents their orbital periods. The third column shows their spin periods and the fourth is the mass of the companion star. The eccentricity is given in the final column. Observational data was taken from the ATNF pulsar catalogue database\citep{Manchester2005}.  }
\begin{tabular}{lllllllll}\hline\hline
PSR Jname&$P_{\rm orb}(\rm day)$&$P_{\rm s}(\rm ms)$&$M_{2}(M_\odot)$&ecc&\\ \hline
J0995-61   &24.578   &     1.99   & 0.22  & 0.11\\
J1623-2631 &191.44   &     1.11   & 0.28  &0.03 \\
J1727-2946 &40.30    &     2.71   & 0.83  & 0.05 \\
J1804-0735 &2.62     &     2.31   & 0.30  & 0.21 \\
J1903+0327 &95.17    &     2.15   & 0.89  & 0.44 \\
J1913+1102 &0.21     &     2.73   & 0.88  & 0.09 \\
J1946+3417 &27.01    &     3.17   & 0.21  & 0.13 \\
J1950+2414 &22.2     &     4.30   & 0.25  & 0.08 \\
J2234+0611 &32.00    &     3.58   & 0.19  & 0.13\\
\hline\hline
\end{tabular}
\label{tab:results}
\end{center}
\end{table}

\begin{figure}
\includegraphics[totalheight=3.5in,width=3.in,angle=-90]{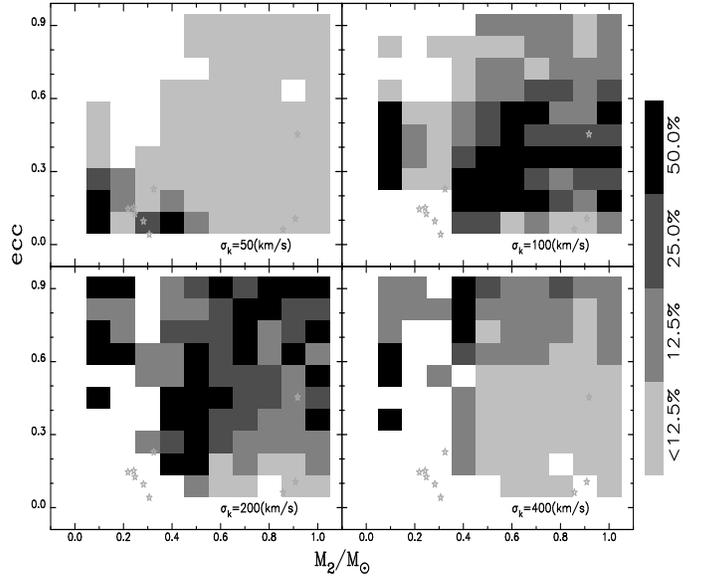}
\caption{Grey-scale map of eccentricity and companion mass for MSP binaries that
have survived post-QN. The various hues of grey correspond to the density of systems,
within 1-1/2, 1/2-1/4, 1/4-1/8, and 1/8-0 of the maximum ${{{\partial^2{N}}\over{\partial {ecc}}{\partial {M_2}}}}$.
Blank regions do not contain any stars.
Stars represent the MSP binaries observed. Observational data was taken
from the Pulsar Catalogue at ATNF\citep{Manchester2005}.}
\label{fig:ecc}
\end{figure}

\begin{figure}
\includegraphics[totalheight=3.5in,width=3.in,angle=-90]{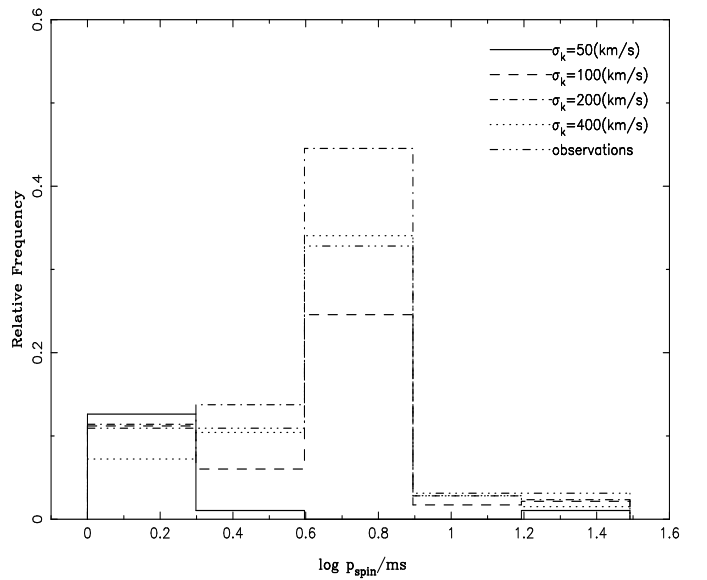}
\caption{Distribution of QN-produced IMSPs¡¯ spin periods with different kick velocities.
The observed spin periods was taken from the Pulsar Catalogue at ATNF \citep{Manchester2005}.}
\label{fig:disps}
\end{figure}

\subsection{IMPSs}
If a QN disrupts an LMXB, then an IMSP will be born. In this paper, IMSPs are assumed to have masses of approximately
 $\sim$1.8 $\rm M_\odot$. As far as current scholarship is concerned, however, it is impossible to measure the true mass of an IMSP.
 This paper also determines the spin period of an IMSP using aforementioned equations (2) and (3). Figure \ref{fig:disps} presents
 the spin periods of IMSPs produced by a QN, each with different kick velocities. According to the ATNF pulsar catalogue database,
there are a total of 64 IMSPs in the galactic field \citep{Manchester2005} and our results are consistent with this observation.
What¡¯s more, results have shown that the higher a QN¡¯s kick velocity is, the more easily IMSPs are produced. We estimate that
the birthrate of QN-produced IMSPs as $\sim 9.5 \times10^{-6}$ yr$^{-1}$
when $\sigma_{\rm k}=50$km s$^{-1}$ and $\sim 1.7\times10^{-4}\rm yr^{-1}$, when $\sigma_{\rm k}=400$km s$^{-1}$.
Despite this, the majority of IMSPs are considered to be the result of the merging of an NS and a massive WD.
The Galactic birthrate of IMSPs produced this way is between $\sim 10^{-4}$  and $10^{-6}$ yr$^{-1}$ \citep{Davies2002,Sun2018}.
Notably, our findings show that QN-produced IMSPs are statistically more significant than those produced by mergers.

The various theories for IMSP formation mentioned above are all related to binary MSPs, including the NS-WD merging theory
\citep{VandenHeuvel1984}, the NS disruption-recycling theory \citep{Camilo1993},
the triple-star system theory \citep{Portegies2011} and the QN formation theory presented in this work.
However, as stated by \cite{Bailes1997} the luminosity of binary MSPs is different from that of others,
such as PSRs J1024-0719, J1744-1134, J2124-3358, and J0711-6830. This suggests that a fraction of IMSPs
may have been produced by another, entirely separate mechanism. In the majority of cases, the space velocity distribution
of IMSPs is wider than that of binary MSPs \citep{Lorimer2004}.
For this reason, IMSPs occupy a larger volume than binary MSPs and are more difficult to detect.
 \cite{Lorimer2008} has suggested that there may even be more IMSPs in the Galaxy if ones takes this selection effect into account.
 In short, scholarship is still far away from fully understanding the origin of all IMSPs.

\section{CONCLUSIONS}
This paper has investigated QNs that occur in LMXBs when the accreting NS¡¯s mass reaches 1.8 $\rm M_\odot$.
If LMXBs are disrupted during this process, then IMSPs are produced. Using population synthesis, this paper
estimated that the Galactic birthrate of QN-produced IMSPs is between $\sim 9.5\times10^{-6}$ and $\sim1.7\times10^{-4}$ $\rm yr^{-1}$.
The uncertainties shown in our experimental model is due to the QN¡¯s kick velocity. Our findings show that
QN-produced IMSPs are statistically more significant than those produced by mergers. The spin period of IMSPs
is also consistent with the distribution given in previous observations. If the LMXBs are not disrupted and
continue to survive after a QN, then they may eventually become MSP binaries with a high eccentricity.

\section*{ACKNOWLEDGEMENTS}
We would like to thank the anonymous referee for their careful reading and constructive criticism.
This work received generous support from the National Natural Science Foundation of China, Project No.
11763007, 11863005, 11803026, and 11503008. We would also like to express our sincere gratitude
to the Tianshan Youth Project of Xinjiang No. 2018Q014.

\bibliography{nn}

\label{lastpage}

\end{document}